\documentclass[useAMS,usenatbib,usegraphicx]{mn2e}

\title[High-significance S--Z measurement: A1914 seen with AMI]
{High-significance Sunyaev--Zel'dovich measurement: Abell~1914 
seen with the Arcminute Microkelvin Imager} 
\author[AMI collaboration]{AMI Collaboration: 
Robert Barker,$^{1}$
Phillip Biddulph,$^{1}$
Dennis Bly,$^{1}$
Roger Boysen,$^{1}$\newauthor
Anthony Brown,$^{1}$
Christopher Clementson,$^{1}$
Michael Crofts,$^{1}$
Thomas Culverhouse,$^{1}$\newauthor
Jaroslaw Czeres,$^{1}$
Roger Dace,$^{1}$
Robert D'Alessandro,$^{1}$
Peter Doherty,$^{1}$\newauthor
Peter Duffett-Smith,$^{1}$
Kenneth Duggan,$^{1}$
John Ely,$^{1}$
Mike Felvus,$^{1}$
William Flynn,$^{1}$\newauthor
J\"orn Geisb\"usch,$^{1}$
Keith Grainge,$^{1}$\thanks{Issuing author -- e-mail:kjbg1@mrao.cam.ac.uk}
William Grainger,$^{2}$
David Hammet,$^{1}$\newauthor
Richard Hills,$^{1}$
Michael Hobson,$^{1}$
Christian Holler,$^{1}$
Roy Jilley,$^{1}$
Michael E. Jones,$^{3}$\newauthor
Takeshi Kaneko,$^{1}$
R\"udiger Kneissl,$^{4}$
Katy Lancaster,$^{5}$
Anthony Lasenby,$^{1}$\newauthor
Phil Marshall,$^{6}$
Francis Newton,$^{1}$
Oliver Norris,$^{1}$
Ian Northrop,$^{1}$\newauthor
Guy Pooley,$^{1}$
Vic Quy,$^{1}$
Richard D. E. Saunders,$^{1}$
Anna Scaife,$^{1}$
Jack Schofield,$^{1}$\newauthor
Paul Scott,$^{1}$
Clive Shaw,$^{1}$
Angela C. Taylor,$^{3}$
David Titterington,$^{1}$
Marko Veli\'c,$^{1}$\newauthor
Elizabeth Waldram,$^{1}$
Simon West,$^{1}$
Brian Wood,$^{1}$
Ghassan Yassin,$^{3}$\newauthor
Jonathan Zwart.$^{1}$\\
$^{1}$Astrophysics Group, Cavendish Laboratory, University of
Cambridge.\\
$^{2}$Columbia University, New York, U.S.A.\\
$^{3}$Astrophysics Group, Denys Wilkinson Building, Univerity of Oxford.\\
$^{4}$Department of Astronomy, Department of Physics, 366 LeConte
Hall, University of California, Berkeley, U.S.A.\\
$^{5}$Astrophysics Group, HH Wills Physics Laboratory, Tyndall
Avenue, Bristol.\\
$^{6}$Kavli Institute for Particle Astrophysics and Cosmology,
Stanford, U.S.A.} 
\begin{document}

\date{Accepted ????. Received ????; in original form ????}

\pagerange{\pageref{firstpage}--\pageref{lastpage}} \pubyear{2005}

\maketitle

\label{firstpage}

\begin{abstract}

We report the first detection of a Sunyaev--Zel'dovich~(S--Z) decrement
with the Arcminute Microkelvin Imager~(AMI). We have made
commissioning observations towards the cluster A1914 and have measured
an integrated flux density of $-8.61$~mJy in a {\it uv}-tapered map with
noise level 0.19~mJy/beam.  We find that the spectrum of the 
decrement, measured in the six channels between 13.5--18~GHz, is consistent
with that expected for a S--Z effect. The sensitivity of the telescope
is consistent with the figures used in our simulations of cluster
surveys with AMI.

\end{abstract}


\section{Introduction}

\begin{figure}
\includegraphics[width=8cm]{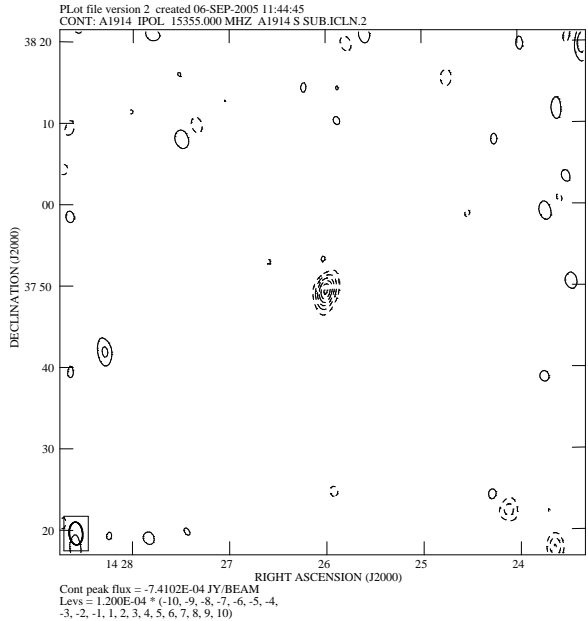}
\caption{{\sc Clean}ed RT map of the S--Z effect in A1914 after source
subtraction. The integrated flux of the S--Z decrement is $-0.76$~mJy
and the noise on the map is 0.06~mJy/beam\label{a1914rt}. The field of
view is chosen to be the same as that in Figures~\ref{a1914nat} and
\ref{a1914tap} for ease of comparison.  In this and subsequent maps:
dashed contours are negative, solid ones positive; the boxed ellipse
in the lower left corner indicates the FWHM of the synthesised beam. }
\end{figure}

The Sunyaev--Zel'dovich effect~\citep{S--Z1972} is a secondary
anisotropy on the Cosmic Microwave Background~(CMB) radiation due to
inverse-Compton scattering of CMB photons from hot plasma in the
gravitational potential of a cluster of galaxies.  The S--Z effect has
been detected by a number of different groups using a variety of
observing techniques~(e.g.  \citealt{birk94}; \citealt{myers};
\citealt{reese}; \citealt{K1999}; \citealt{P1999}; \citealt{H1997};
\citealt{G1993}; \citealt{L2005}; \citealt{U2004}.  See
e.g. \citealt{B1999} or \citealt{CHR2002} for a full review).  For
example, in the past, the Cambridge group used the five antennas of
the Ryle Telescope (RT) in a compact configuration to produce S--Z
maps such as that shown in Figure~\ref{a1914rt}~\citep{J2005}.  We
detected a S--Z decrement towards the cluster A1914 with an
integrated flux of $-0.76$~mJy and a map noise of 0.06~mJy/beam from a
228-hour observation.  This long integration time was required because
the RT has relatively poor system temperature, has bandwidth of only
350~MHz and because the telescope baselines were long compared to the
optimum required to match to the angular size of the cluster, with the
result that most of the S--Z signal is resolved out.

The S--Z effect has the unique property that its surface brightness is
independent of redshift. This makes it ideal as a means of surveying
for galaxy clusters back to their epoch of formation. Several
dedicated instruments are now being built to conduct such S--Z
surveys~(see e.g. \citealt{S--ZA,AMIBA,SPT,ACT,K2001}).  At Cambridge, the
Arcminute Microkelvin Imager~(AMI) is being commissioned. This
instrument has been designed to make high-speed, sensitive surveys for
S--Z decrements by using many small dishes packed together in a compact
array with a low noise, high-bandwidth back-end system and
complemented by an array of much larger dishes designed to detect and
subtract compact foreground radio sources.  In this paper we present a
commissioning observation with AMI towards A1914 which has previously
been observed in S--Z~\citep{G2001,J2005} and in
X-ray~(see e.g. \citealt{bcs1-98}; \citealt{noras1-00}; \citealt{I2002}).

\section{The Arcminute Microkelvin Imager}

AMI~\citep{J2002} comprises two synthesis arrays, one of ten 3.7-m
antennas (Small Array) and one of eight 13-m antennas (Large Array),
both sited at Lord's Bridge, Cambridge. The Large Array is an upgraded
version of the RT, with the three outlying antennas now
moved into a compact configuration near the five which we previously
used for S--Z work. The telescope observes in the band
12--18~GHz with cryostatically cooled NRAO indium-phosphide front-end
amplifiers. The overall system temperature is approximately 25~K.
The radio frequency is mixed with a 24-GHz
local oscillator, downconverting to an intermediate frequency~(IF) band
of 6--12~GHz. Amplification, equalisation, path compensation and
automatic gain control are then applied to the IF signal.
The correlator is an analogue Fourier transform spectrometer with 16
correlations formed for each baseline at path delays spaced by
25~mm. In addition, both `$+$' and `$-$' correlations are formed by
use of $0{\degr}$ and $180{\degr}$ hybrids respectively. From these,
eight 0.75-GHz channels are synthesised.

\section{Observations}

The new observations presented here are commissioning runs with just
eight antennas of the Small Array and only the upper six frequency
channels (giving a total of 4.5~GHz of bandwidth). We observed
Abell~1914 for a total of 34~hours between 2005 August 15 and 2005
August 29 with a pointing centre of 14~26~02.15
+37~50~05.8~(J2000). The observations were phase and amplitude
calibrated on 3C286. We take the flux density of 3C286 as
3.48~Jy~\citep{B1977} in $I+Q$ at 15~GHz with a spectrum of
$\alpha=0.733$ ($S\propto \nu^{-\alpha}$); a correction factor of 1.05
has been assumed to account for the $\approx~12\%$ polarisation at
position angle $33\degr$ of the source. Specialist AMI data reduction
software (the {\sc Reduce} package) was used to flag the data for
telescope pointing errors, to excise interference, to apply the
calibration and to weight the data. Reduced data were then transferred
to {\sc Aips} for further analysis.

\section{Results}

\begin{figure}
\includegraphics[width=8cm]{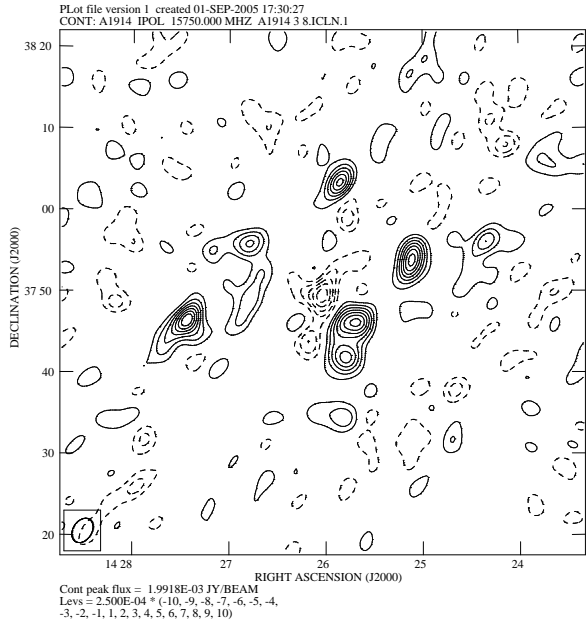}
\caption{Naturally weighted, {\sc Clean}ed AMI map of the A1914 field
before source subtraction. The noise on the map is
0.15~mJy/beam\label{a1914nat}.}
\end{figure}

\begin{figure}
\includegraphics[width=8cm]{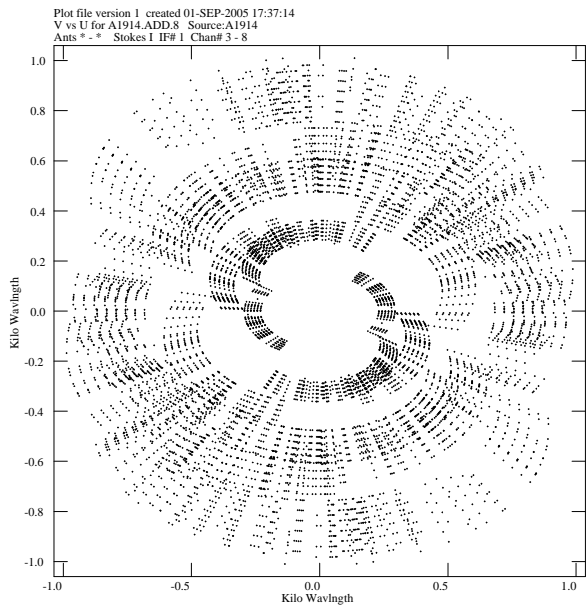}
\caption{Aperture plane coverage for the six channels used.\label{a1914uvcov}}
\end{figure}

\begin{figure}
\includegraphics[height=8cm,angle=270]{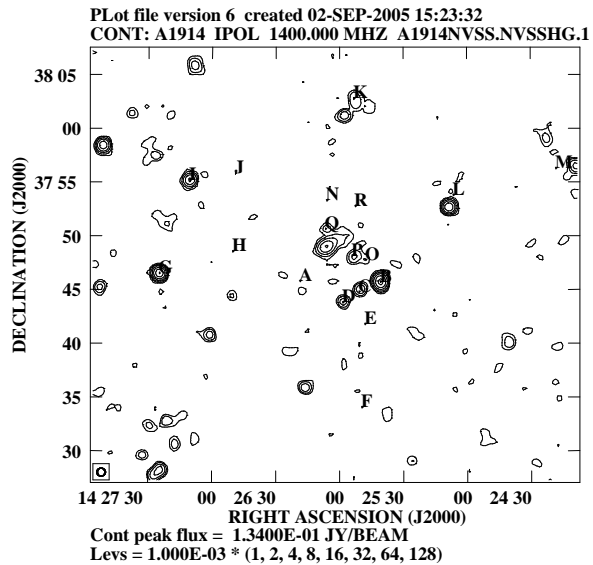}
\caption{NVSS map of the A1914 field indicating the positions of the 18
sources subtracted from the AMI data.\label{a1914nvss}}
\end{figure}

A naturally weighted map of the A1914 field is shown in
Figure~\ref{a1914nat} and a plot showing the coverage of the aperture
plane in Figure~\ref{a1914uvcov}. The negative feature at the centre
of the {\sc Clean}ed map is the S--Z effect from the cluster, but it
is confused by several radio sources which are also visible and whose
effects must be subtracted. The brighter sources were mapped
individually with the RT at the same frequency to determine their
positions and fluxes. Contributions from sources close to the cluster
centre were estimated from the 19-day RT observations (Section 1). Once
these sources were removed, the AMI data were remapped and the fluxes
of any remaining faint sources estimated and removed from the
visibilities, a method we have found to be successful with previous RT
S--Z observations~(see e.g. \citealt{G1996}). In all, 18 sources were
subtracted from the visibility data as described in
Table~\ref{sources}. The positions of the subtracted sources are shown
on an NVSS~\citep{C1998} map in Figure~\ref{a1914nvss}.

\begin{table}
\caption{Sources subtracted from the S--Z map. The flux densities are
those subtracted from the visibilities, i.e. not corrected for the
primary beam.\label{sources}\newline}
\begin{tabular}{lllllllllll}
Source  & RA (J2000)  & Dec. (J2000)	& Flux density (mJy) & \\
\hline%--------------------------------------------------------
A	& 14 26 18.60 & +37 45 51.0	& 0.70	   & \\ % M
B	& 14 25 40.70 & +37 45 46.4	& 1.90	   & \\ % A1
C	& 14 25 50.80 & +37 44 51.0	& 0.65	   & \\ % A2
D	& 14 25 58.40 & +37 43 51.0	& 0.65	   & \\ % A3
E	& 14 25 48.00 & +37 41 51.0	& 1.4	   & \\ % J
F	& 14 25 49.50 & +37 34 06.0	& 1.02	   & \\ % O
G	& 14 27 24.82 & +37 46 33.1	& 1.91	   & \\ % B
H	& 14 26 50.20 & +37 48 35.0	& 0.82	   & \\ % N
I	& 14 27 10.53 & +37 55 14.0	& 0.32	   & \\ % C
J	& 14 26 49.10 & +37 55 50.2	& 0.91	   & \\ % L
K	& 14 25 53.30 & +38 02 51.0	& 1.6	   & \\ % E1
L	& 14 25 06.40 & +37 53 50.0	& 2.1	   & \\ % D
M	& 14 24 18.00 & +37 56 18.0	& 1.05	   & \\ % K
N	& 14 26 06.00 & +37 53 21.0	& 0.84	   & \\ % P
O	& 14 25 47.60 & +37 47 48.8	& 0.54	   & \\ % F
P	& 14 25 54.00 & +37 48 13.1	& 0.24	   & \\ % G
Q	& 14 26 06.50 & +37 50 41.6	& 0.20	   & \\ % I
R	& 14 25 52.60 & +37 52 49.0	& 0.52	   & \\ % H
\hline%--------------------------------------------------------

\end{tabular}
\end{table}

\begin{figure}
\includegraphics[width=8cm]{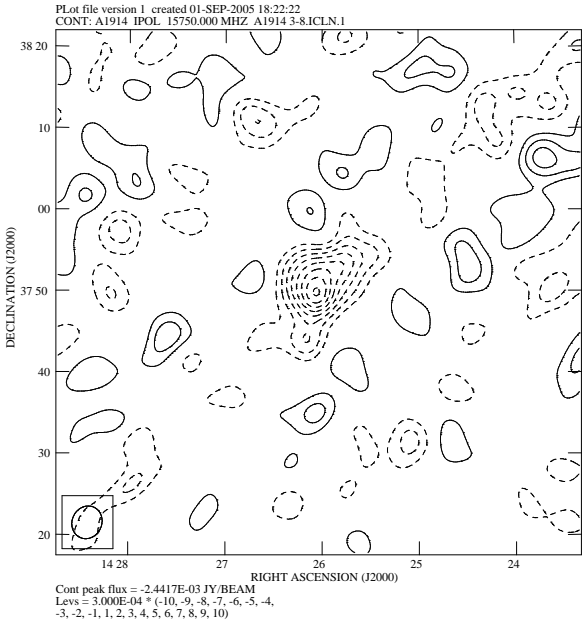}
\caption{{\sc Clean}ed map of the S--Z effect in A1914 after source
subtraction and with a {\it uv}-taper of $500~\lambda$, resulting in a
resolution of $247\times216\arcsec$. The noise on
the map is 0.19~mJy/beam\label{a1914tap}. The S--Z effect is clearly resolved
and has an integrated flux density of $-8.61$~mJy.}
\end{figure}

\begin{figure}
\includegraphics[height=8cm,angle=270]{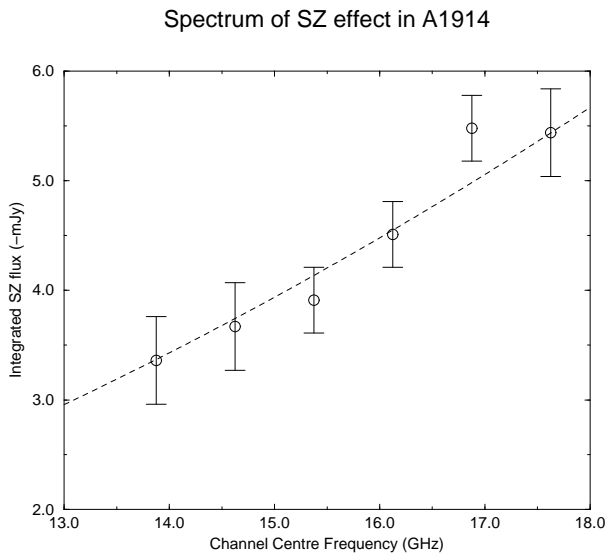}
\caption{Spectrum of the S--Z effect in A1914. \label{a1914spec} The
integrated flux is found over a 56 square arcminute region at the
cluster centre. The magnitude of the negative S--Z flux is shown for
convenience. The error bars are the $1\sigma$ errors from the channel
maps which are dominated by source confusion; errors between different
channels are therefore not independent. The dashed line shows a
thermal spectrum ($\alpha = -2$ in the Rayleigh-Jeans region) 
constrained to pass through the 13.875~GHz point.}
\end{figure}

The source-subtracted data were then remapped with a {\it uv}-taper of
$500~\lambda$ to increase the sensitivity to the S--Z effect and are
shown in Figure~\ref{a1914tap}. The integrated S--Z flux in the map is
$-8.61$~mJy and the noise on the map is 0.19~mJy/beam. It is also possible
to make maps of individual frequency channels and use these to measure
the frequency spectrum of the S--Z effect. Since the S--Z effect is
extended and the aperture plane coverage for each channel is different
(and hence samples different angular scales), it is necessary to
measure the flux from each of the channels over the same solid
angle. Figure~\ref{a1914spec} shows the integrated S--Z flux density
over a 56-square-arcminute region at the cluster centre against
channel centre frequency.

\section{Discussion}

The S--Z effect is detected with extremely high significance
(Figure~\ref{a1914tap}) and extended structure is clearly visible.
Comparing the map from 34~hours of AMI observation with the previous
RT map demonstrates the potential of AMI as a cluster survey
instrument.  Maps of short observations that are limited by thermal
noise show that the Small Array sensitivity per channel is
approximately $350~\rm mJy~s^{-1/2}$, which is consistent with that
assumed in our simulations~\citep{K2001}. Calculations based on the
9C~survey source counts~\citep{W2003} show that the AMI maps presented here
are limited by source confusion~\citep{S1957} rather than thermal
noise.  This and the fact that we have had to remove 18 point sources
from our data emphasise the importance of source subtraction, which
for AMI will be provided by the Large Array.

Figure~\ref{a1914spec} shows that the spectrum of the 
decrement measured by AMI is consistent with an S--Z spectrum, which
differs from a thermal spectrum by less than $0.5\%$ between
13.5 and 18~GHz~\citep{CL1998}.

In order to make a preliminary comparison of the AMI data on A1914
with X-ray observations, we have fitted an isothermal spherical
$\beta$-model, assuming a concordance cosmology
($H_0=72 \, \rm{km \, s^{-1} \, Mpc^{-1}}$, $\Omega_M =
0.3$, $\Omega_\Lambda=0.7$), to Chandra ACIS-I data using a
temperature $T=8.41 \, \rm{keV}$~\citep{I2002}, redshift $z=0.1712$, a
count rate calculated from NORAS~PSPC and $n_H$ using the HEASARC webPIMMS
tool with a Raymond Smith profile~\citep{noras1-00}.  We find best fit
parameters of $\beta=0.731$, central cluster density
$n_0=0.01627 \, \rm{cm^{-3}}$ and core radius $\theta_c = 51.0\arcsec$.
Using this parameterisation of the intracluster gas, we produce a
simulated S--Z observation of the cluster following~\citet{G2002}. We
map and {\sc Clean} this simulated data set in the same manner as the
real data and find that the predicted integrated S--Z flux from the
X-ray model is $-7.35$~mJy. This is in good agreement with the value
of $-8.61$~mJy given the assumptions implicit in the model (e.g. that
the line of sight depth through the cluster equals the size of the
cluster projected on the sky). A more sophisticated simulation of the
intracluster gas is beyond the scope of this paper.

The AMI data discussed in this paper come from a
commissioning run; further system improvements, such as
weighting for system temperature variations through synchronous measurement
of injected noise, are underway.

\section*{Acknowledgment }

We thank PPARC for support for AMI and its operation.

\label{lastpage}

\end{document}